\documentclass{PoS}

\newcommand{\beq}{\begin{equation}}
\newcommand{\eeq}{\end{equation}}

\newcommand{\bea}{\begin{eqnarray}}
\newcommand{\eea}{\end{eqnarray}}
\newcommand{\ba}{\begin{array}}
\newcommand{\ea}{\end{array}}

\title{Baryon resonances at large $N_c$, \\or Quark Nuclear Physics}

\ShortTitle{Baryon resonances}

\author{\speaker{Dmitri Diakonov}\\
        Petersburg Nuclear Physics Institute, Gatchina 188000, St. Petersburg, Russia\\
        Academic University, 195220, St. Petersburg, Russia \\
        E-mail: \email{dmitri.diakonov@gmail.com}}

\author{Victor Petrov\\
        Petersburg Nuclear Physics Institute, Gatchina 188000, St. Petersburg, Russia\\
        E-mail: \email{victorp@thd.pnpi.spb.ru}}

\author{Alexey A. Vladimirov\\
        Bochum University, D-44780, Bochum, Germany\\
        E-mail: \email{vladimirov.aleksey@googlemail.com}}

\abstract{We suggest a new point of view according to which baryon resonances can be
understood as collective excitations about intrinsic one-quark excitations
in a mean field of definite symmetry. This approach is justified in the limit of 
large number of colours $N_c$, and is similar to the physics of large-$A$ nuclei, 
hence ``quark nuclear physics''.  Although in the real world $N_c$ is only three, 
we obtain a good agreement with the observed resonance spectrum of light baryons 
up to 2 GeV, and of lowest charmed baryon multiplets. A by-product of the scheme 
is the prediction of a new exotic charmed (and bottom) baryons that may be stable
against strong decays.}

\FullConference{Sixth International Conference on Quarks and Nuclear Physics,\\
		April 16-20, 2012\\
		Ecole Polytechnique, Palaiseau, Paris}

\begin{document}

\section{Introduction}

Baryons from the large-$N_c$ point of view have been studied by quite a number of people who
derived many algebraic relations between baryon observables, following from the requirement
that physical quantities should have ``natural'' behaviour in $N_c$, see
{\it e.g.}~\cite{Manohar:2001cr,Goity:2005fj,Cohen:2005uv} and references therein. In this report,
we suggest a simple physical picture that results in those relations, and derive some 
new ones. A longer paper has been recently published~\cite{Diakonov:2012zz}, and a detailed report 
is currently in preparation.

If the number of colours $N_c$ is large the $N_c$ quarks constituting a baryon can be considered
in a mean (non-fluctuating) field that is stable as $N_c\!\to\!\infty$~\cite{Witten:1979kh}.
At the microscopic level quarks experience colour interactions, however gluon field fluctuations
are not suppressed if $N_c$ is large; the mean field can be only `colourless'. An example
how originally colour interactions transform mathematically into interactions of dynamically massive 
quarks with mesonic fields is provided by the instanton liquid model, see {\it e.g.}~\cite{Diakonov:2002fq}.
A non-fluctuating confining chiral bag model gives another example of a `colourless' mean field.
A modern example is the 5- or 6-dimensional `gravitational', plus mesonic, background field
in the holographic QCD models. They also need large $N_c$ for a justification.

The advantage of the large-$N_c$ approach is that at large $N_c$ baryon physics simplifies
considerably. It becomes possible to take into full account the important relativistic
and field-theoretic effects that are often ignored. Baryons are not just three (or $N_c$)
quarks but contain additional quark-antiquark pairs, as it is well known experimentally.
The number of antiquarks in baryons is, theoretically, also proportional to $N_c$~\cite{Diakonov:2008hc},
which means that antiquarks cannot be obtained from adding one or two mesons to a baryon:
one needs ${\cal O}(N_c)$ mesons to explain ${\cal O}(N_c)$ antiquarks, implying in fact
a classical mesonic field.

If the mean field at large $N_c$ is a reality, one can consider quarks in that mean
field -- similarly to nucleons in the mean field of large-$A$ nuclei. Generally, quarks
obey the Dirac equation in the background field, that may be in fact non-local. All intrinsic
quark Dirac levels in the mean field are stable in $N_c$. All negative-energy levels should
be filled in by $N_c$ quarks in the antisymmetric state in colour, corresponding to the zero baryon
number state. Filling in the lowest positive-energy level by $N_c$ `valence' quarks makes
a baryon, see Fig.~1, left. Exciting higher quark levels or making particle-hole excitations 
produces baryon resonances~\cite{Diakonov:2008rd,Diakonov:2009kw}, see Fig.~1, middle. 
The baryon mass is ${\cal O}(N_c)$, and the excitation energy is ${\cal O}(1)$. When one excites 
one quark the change of the mean field is ${\cal O}(1/N_c)$ that can be neglected to the first 
approximation.

Moreover, if one replaces one light ($u,d$ or $s$) quark in light baryons by a heavy
($c,b$) one, as in charmed or bottom baryons, the change in the mean field is also
${\cal O}(1/N_c)$. Therefore, the spectrum of heavy baryons is directly related to
that of light baryons~\cite{Diakonov:2009kw,Diakonov:2010zz}. In light baryons made of 
$u,d,s$ quarks, we first take the chiral limit of zero quark masses. It implies that 
all resonances appear as degenerate $SU(3)_{\rm flav}$ multiplets that are later split 
by nonzero $m_s$.

Next, we argue that the mean field in baryons has a definite symmetry, namely it breaks 
spontaneously the symmetry under separate $SU(3)_{\rm flav}$ and $SO(3)_{\rm space}$ 
rotations but does not change under simultaneous $SU(2)_{\rm iso+space}$ rotations in 
ordinary space and a compensating rotation in isospace~\cite{Diakonov:2008rd,Diakonov:2009kw}.
Similarly, in nuclear physics the $SO(3)_{\rm space}$ symmetry is broken spontaneously
in the ground state since most of the heavy nuclei are elliptical-, not spherical-symmetric.

Spontaneous breaking of $SU(3)_{\rm flav}$ means, in particular, that one-quark levels
for $u,d$ quarks on the one hand, and for $s$ quarks on the other, are totally, 100\% different
even in the chiral limit $m_s\to 0$. The latter levels are characterized by total angular
momentum ${\bf J}={\bf L}+{\bf S}$, and parity, whereas the former are characterized by the
grand spin ${\bf K}={\bf J}+{\bf T}$ where ${\bf T}$ is the isospin, and parity. 

The original $SU(3)_{\rm flav}\times SO(3)_{\rm space}$ symmetry is restored by the (quantized)
rotation of the mean field in flavour and ordinary spaces. It implies that each intrinsic 
quark state, be it the ground state or a one-quark excitation in the Dirac spectrum, 
generates a band of $SU(3)$ multiplets appearing as collective rotational excitations of 
a given intrinsic state. The quantum numbers of those multiplets, their total number and 
the ${\cal O}(1/N_c)$ splittings between them are unequivocally dictated by the symmetry 
of the mean field, see Fig.~1, right. Assuming the $SU(2)_{\rm iso+space}$ symmetry of 
the mean field, we obtain exactly the spin and flavour multiplets that are observed in Nature.

All properties of baryon resonances belonging to one band associated with a given one-quark 
excitation are related by symmetry. For example, the splittings inside several $SU(3)$ 
multiplets as due to the nonzero $m_s$, are related to each other: these relations are 
satisfied with high accuracy, in some cases better than the Gell-Mann--Okubo relations 
for separate $SU(3)$ multiplets.

At this time we do not consider any specific dynamical model but concentrate mainly on symmetry. 
A concrete dynamical model would say what is the intrinsic relativistic quark spectrum in baryons. 
Instead of calculating the intrinsic Dirac spectrum of quarks from a model, we extract it 
from the experimentally known baryon spectrum by interpreting baryon resonances as collective 
excitations about the ground state and about the one-quark transitions. However, we show 
in Ref.~\cite{Diakonov:2012zz} that the needed intrinsic quark spectrum can be obtained from 
a natural choice of the mean field satisfying the $SU(2)_{\rm iso+space}$ symmetry.

In summary, we show that it is possible to obtain a realistic spectrum of baryon
resonances up to 2 GeV, starting from the large-$N_c$ limit. The lowest $SU(3)$ multiplets
of charmed baryons also fit nicely into this universal picture~\cite{Diakonov:2009kw,Diakonov:2010zz}.

\begin{figure}
\includegraphics[width=.34\textwidth]{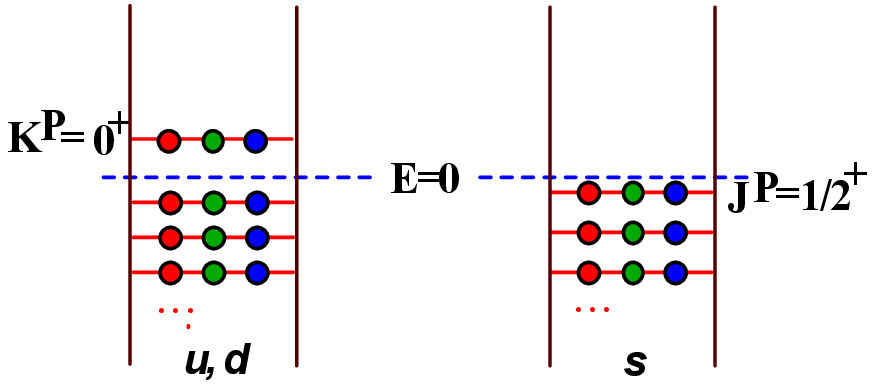}\hspace{0.65cm}
\includegraphics[width=.34\textwidth]{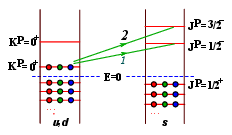}\hspace{0.65cm}
\includegraphics[width=.19\textwidth]{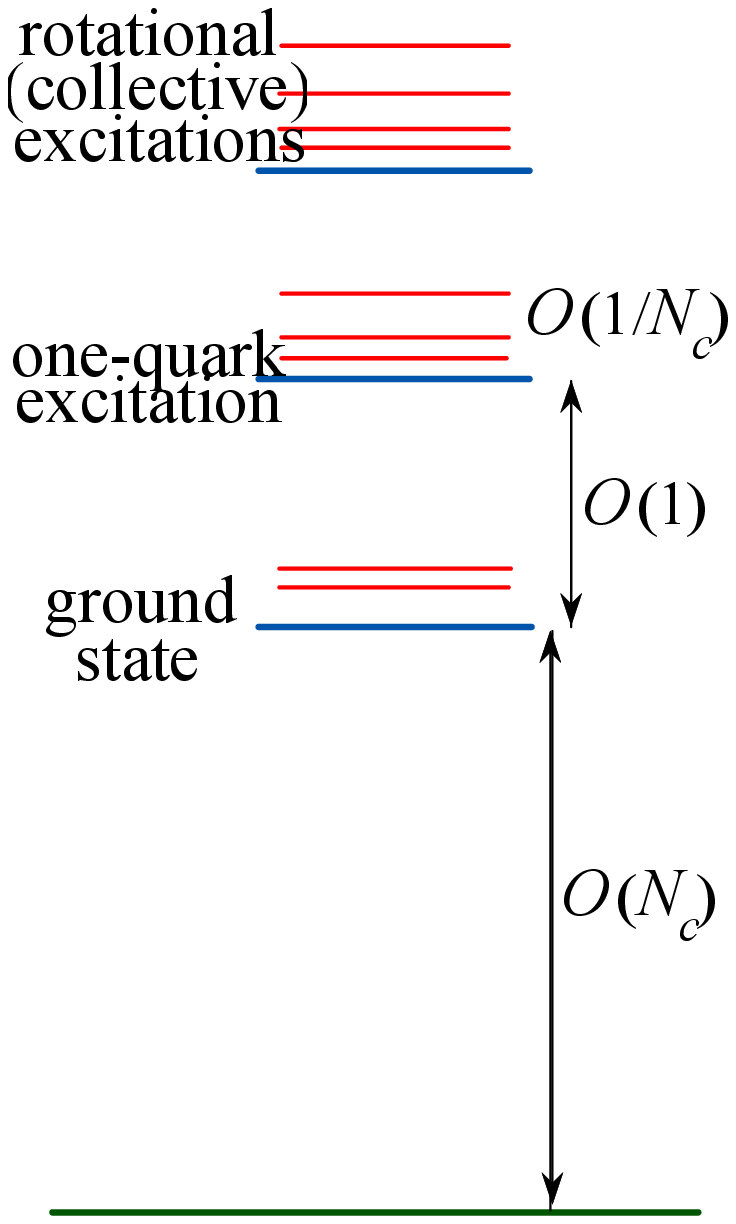}
\caption{{\it Left}: Filling in $u,d$- and, separately, $s$-quark levels for the ground-state 
light baryons. {\it Middle}: Examples of one-quark excitations. 
{\it Right}: Hierarchy of scales in large-$N_c$ baryons.}
\label{fig1}
\end{figure}

\section{Baryons made of $u,d,s$ quarks}

The ground-state baryons are obtained from the level-filling scheme shown in Fig.~1, left.
The quantization of the rotations (restoring the original flavour and space symmetries) of the
$K^P=0^+$ valence level gives rise to the lowest baryons multiplets, $({\bf 8}, 1/2^+)$
and $({\bf 10}, 3/2^+)$. 

The singlets, $({\bf 1}, 1/2^-)$ and $({\bf 1}, 3/2^-)$ (the $\Lambda$ hyperons), are obtained
from ``Gamov--Teller'' excitations of the $s$-quark levels, see Fig.~1, middle. In principle, these
two transitions entail rotational bands, however the rotational splitting of higher rotational 
excitations are not ${\cal O}(1/N_c)$ but ${\cal O}(1)$~\cite{Diakonov:2010zz}, therefore all states
except the lowest-mass singlets must be ignored in this approach. 

The parity-plus resonances are obtained from the $u,d$ one-quark excitations $0^+\to 0^+$,
$0^+\to 1^+$ and $0^+\to 2^+$, whereas parity-minus resonances can be obtained from
the $u,d$ transitions $0^+\to 0^-$, $0^+\to 1^-$ and $0^+\to 2^-$, see ~\cite{Diakonov:2012zz}.
All known baryon resonances below 2 GeV can be nicely fit into the rotational states stemming
from those one-quark excitations. There are no resonances from the PDG tables that are left 
unaccounted for, and there are no extra states from the theory side, except the $\Delta(3/2^+)$
resonance coming from the rotational excitation on top of the $0^+\to 1^+$ transition. It
is therefore our prediction.

\begin{figure}
\includegraphics[width=.31\textwidth]{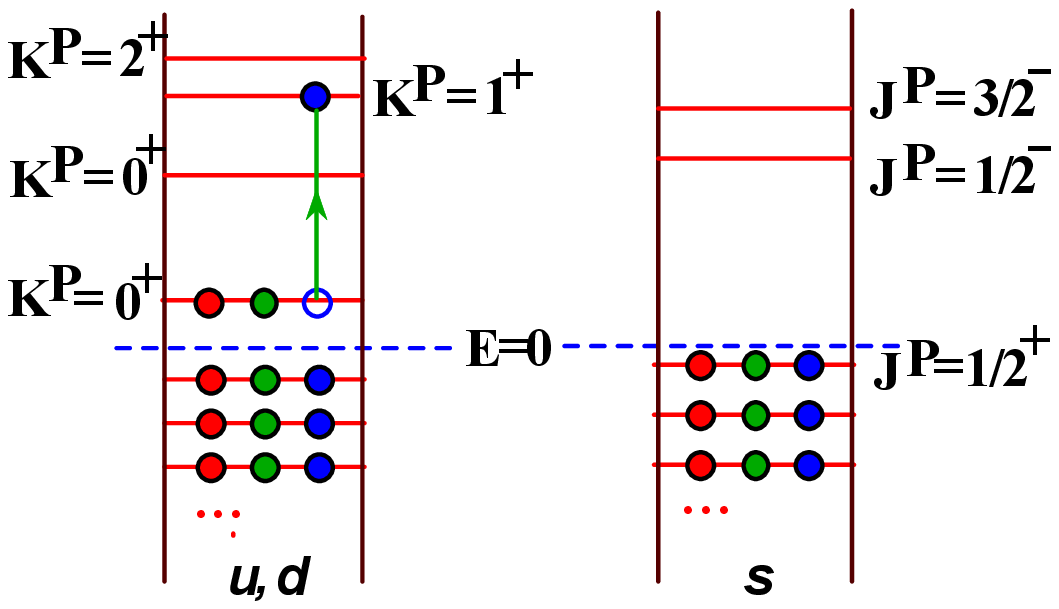}\hspace{0.35cm}
\includegraphics[width=.31\textwidth]{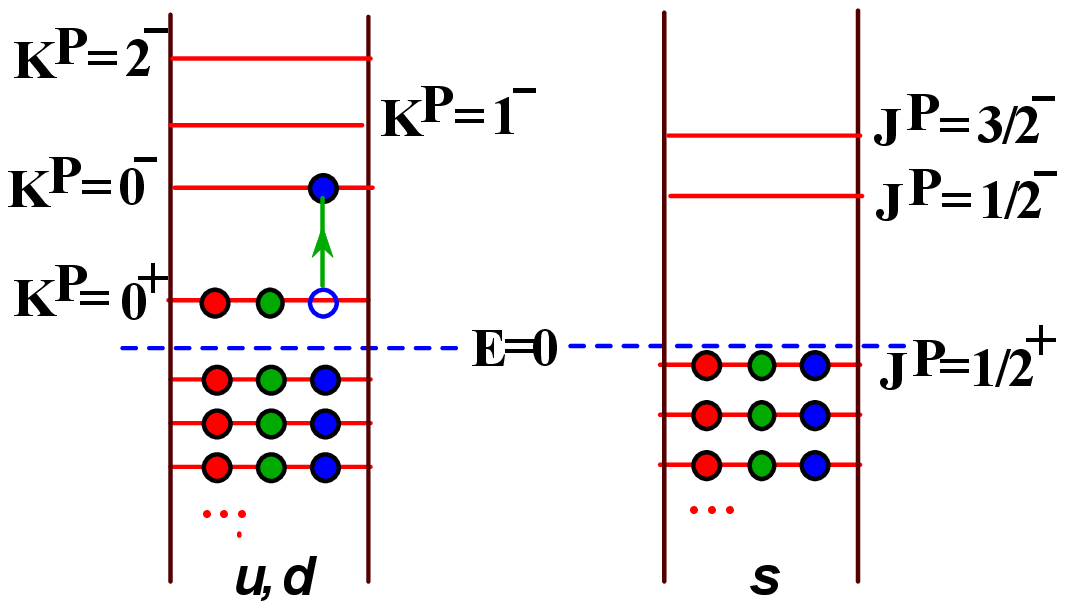}\hspace{0.35cm}
\includegraphics[width=.31\textwidth]{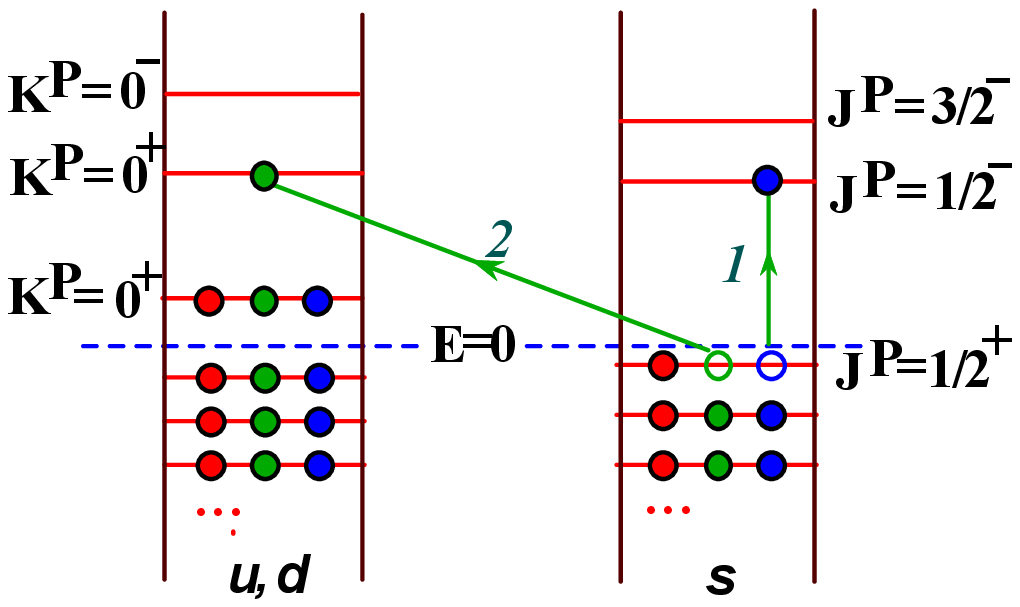}
\caption{{\it Left}: The origin of parity-plus light resonances. {\it Middle}: The origin of parity-minus
light baryons. {\it Right}: Pentaquarks are particle-hole excitations of a negative-energy $s$ quark.
The exotic parity-plus $\Theta^+$ (recently seen again at JLab~\cite{Amaryan:2011qc}) corresponds to a 
non-diagonal or `Gamov--Teller' transition {\it 2}.}
\label{fig2}
\end{figure}

\section{Splittings inside $SU(3)$ multiplets}

Properties of baryon resonances, such as the couplings, partial decay widths, mass splittings, {\it etc.}, 
arising from one rotational band about a given one-quark excitation are related to each other by symmetry. 
Typically, the observables for resonances  from {\em different} multiplets are expressed through a small 
number of constants, which leads to numerical relations between the observables, that can be tested 
against the data.

For example, in the linear order in $m_s$ the splittings inside an octet are determined by two constants
$\mu_{1,2}^{(8)}$ while those in a decuplet are determined by only one constant $\mu^{(10)}$. We write the masses
of individual members of the octets,
$$
{\cal M}_{N}=M_8-\frac{7}{4}\mu^{(8)}_1-\mu^{(8)}_2, \quad  {\cal M}_{\Lambda}=M_8-\mu^{(8)}_1, 
\quad {\cal M}_{\Sigma}=M_8+\mu^{(8)}_1 \quad {\cal M}_{\Xi}
=M_8+\frac{3}{4}\mu^{(8)}_1+\mu^{(8)}_2,
$$
and the masses of the members of the decuplets,
$$
{\cal M}_{\Delta}=M_{10}-\mu^{(10)}, \qquad  {\cal M}_{\Sigma}=M_{10}, \qquad {\cal M}_{\Xi}=M_{10}+\mu^{(10)} \qquad {\cal M}_{\Omega}
=M_{10}+2\mu^{(10)}.
$$    
From this parametrization one gets automatically the Gell-Mann--Okubo relations for the octets and decuplets
separately. However, the symmetry breaking pattern $SU(3)_{\rm flav}\times SO(3)_{\rm space}\to SU(2)_{\rm iso+space}$
that we advocate enables one to relate the splitting parameters related to {\em different} multiplets
with various spins. In particular, we obtain the following new relations for the different multiplets 
stemming from the $0\to 1$ and $0\to 2$  one-quark transitions, respectively,
$$
7\mu^{(10)}\left(\frac{1}{2}\right)+3\mu^{(8)}_2\left(\frac{3}{2}\right)=10\mu^{(10)}\left(\frac{3}{2}\right),
$$
$$
5\mu^{(8)}_2\left(\frac{5}{2}\right)+11\mu^{(10)}\left(\frac{3}{2}\right)=16\mu^{(10)}\left(\frac{5}{2}\right)
$$
(in the parentheses we indicate the spin of the multiplet). Fitting the masses of individual multiplet members
by the parameters $\mu$ one can find them for various multiplets and then plug into the above relations.
It turns out that these very nontrivial relations are satisfied to an accuracy of 1-2\%! It gives a strong
support to the scheme suggested. Another support comes from the fact that the ${\cal O}(1/N_c)$ 
splittings between the {\it centers} of the $SU(3)$ multiplets are expressed through two moments of inertia,
and phenomenologically many splittings can be described, to a fair accuracy, with the same moments of inertia.
Surprisingly, approximately the same moments of inertia are needed to explain the splittings in heavy baryons.

\section{Heavy baryons}

If one of the light quarks in a light baryon is replaced by a heavy $b$ or $c$ quark,
there are still $N_c\!-\!1$ light quarks left. At large $N_c$, they form {\em the same} mean field as
in light baryons, with the same sequence of Dirac levels, up to $1/N_c$ corrections. The heavy quark
contributes to the mean $SU(3)_{\rm flav}$ symmetric field but it is a $1/N_c$ correction, too. It means
that at large $N_c$ one can {\em predict} the spectrum of the $Qq\ldots q$ (and $Qq\ldots qq\bar q$) baryons from
the spectrum of light baryons. 

The filling of Dirac levels for the ground-state $c$ (or $b$) baryon is shown in Fig.~3, left:
there is a hole in the $0^+$ shell for $u,d$ quarks as there are only $N_c-1$ quarks there, in an antisymmetric
state in colour. Adding the heavy quark makes the full state `colourless'.

\begin{figure}[h]
\begin{center}
\includegraphics[width=.33\textwidth]{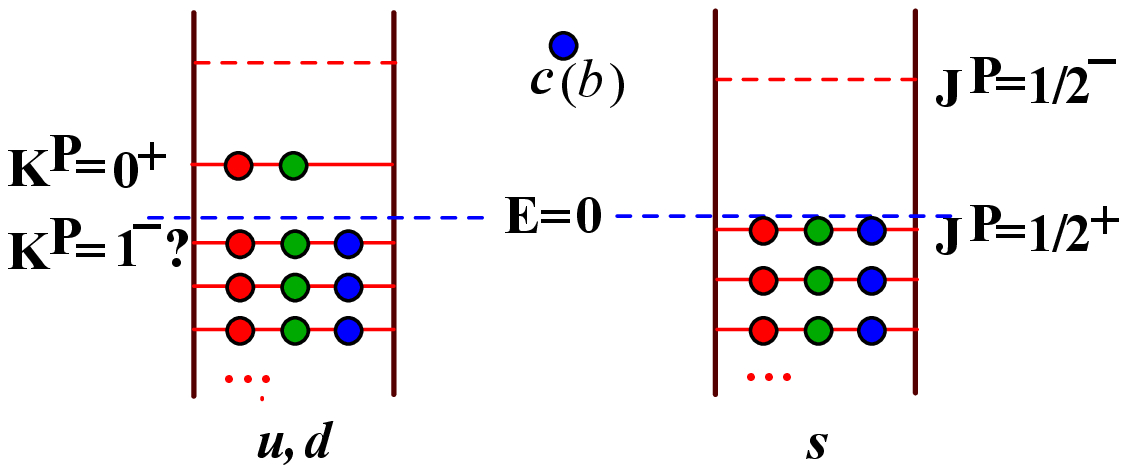}\hspace{0.3cm}
\includegraphics[width=.33\textwidth]{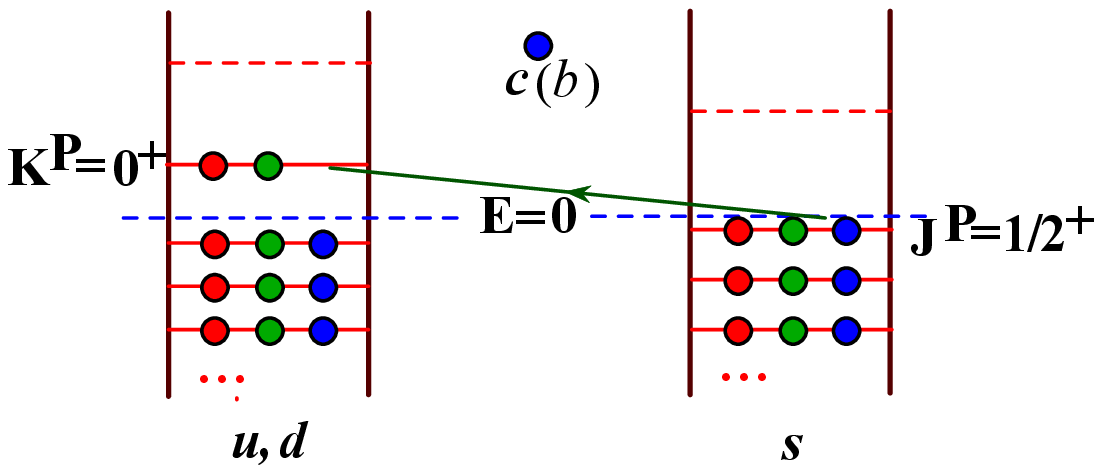}\hspace{0.3cm}
\includegraphics[width=.28\textwidth]{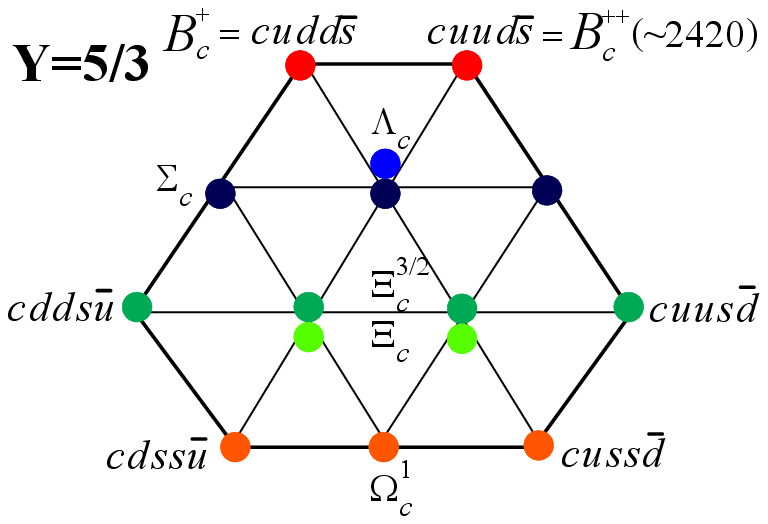}
\end{center}
\caption{{\it Left}: Ground-state charmed (or bottom) baryon with a hole in the valence $u,d$ level.
{\it Middle}: A particle-hole excitation of the $s$-quark from the negative-energy level. {\it Right}:
The corresponding ${\bf\overline{15}}$-plet of heavy pentaquarks. The {\em Beta} baryons in the upper
line may be stable against strong decays.}
\label{fig3}
\end{figure}

As in the case of light baryons, the filling scheme by itself does not tell us what are the quantum numbers
of the state: they arise from quantizing the $SU(3)_{\rm flav}$ and $SO(3)_{\rm space}$ rotations of the given
filling scheme. The quantization of the rotations about the ground state gives three $SU(3)$ multiplets:
$({\bf \bar 3}, 1/2^+), ({\bf 6}, 1/2^+)$ and $({\bf 6}, 3/2^+)$. These are exactly the lightest charmed
multiplets observed in Nature. The centers of the last two multiplets are split by $1/N_c^2$ and $1/m_c$ 
corrections, and indeed the splitting is small (67 MeV), whereas the splitting between the centers of the ${\bf 6}$'s
and the ${\bf \bar 3}$ is ${\cal O}(1/N_c)$ and is in good accordance with the splitting between {\em light}
rotational states, thus supporting the main idea~\cite{Diakonov:2009kw,Diakonov:2010zz}.  

Finally, the hole in the ground-state $0^+$ shell can be filled in either by exciting a $u,d$ quark 
from lower lying shells, or by non-diagonal `Gamov-Teller' transition from the highest filled $s$ quark shell,
see Fig.~3, middle. In the second case the corresponding charmed baryon resonance will be a pentaquark exotic,
actually belonging to a ${\bf \overline{15}}$-plet from the $SU(3)$ point of view, Fig.~3, right. 

The masses of the exotic ``Beta baryons'', ${\cal B}_c^{++}=cuud\bar s,\;{\cal B}_c^+=cudd\bar s$, and
${\cal B}_b^+=buud\bar s$, ${\cal B}_b^0=budd\bar s$ have been estimated in Ref.~\cite{Diakonov:2009kw,Diakonov:2010zz}
to be quite light, $m_{\cal B}\approx 2420\,{\rm MeV}$, and hence they may be well stable against
strong decays! See Ref.~\cite{Diakonov:2010zz} for the discussion of the ${\bf \overline{15}}$-plet of
the exotic charmed (or bottom) pentaquarks and of the possibilities of their observation at LHC
and b-factories.
%\vskip 0.5true cm

%\underline{\bf Conclusions}. 
\section{Conclusions}

We have presented a new unified picture of baryon resonances, light and heavy, 
that is in fact quite similar to that of nuclei at large $A$ whose r\^ole is played by large $N_c$. 
While $N_c$ is only three in the real world, we see that the imprint of the large-$N_c$ picture is 
clearly visible. A by-product of this study is the prediction of a relatively light exotic charmed pentaquark, 
 the {\it Beta}.

\end{document}